\documentclass{ws-procs975x65}

\usepackage{xcolor}
\definecolor{darkgreen}{hsb}{.333,1,.5}
\definecolor{darkblue}{hsb}{.667,1,.5}
\usepackage{hyperref}
\hypersetup{colorlinks=true,
        linkcolor=darkblue,
        linktoc=all,
        citecolor=darkgreen,
        urlcolor=blue}
\usepackage[all]{hypcap}	

\def\beq{\begin{equation}}
\def\eeq{\end{equation}}

\begin{document}

\newcommand{\schwJoopathvisfig}{
    \begin{figure}[t!]
    \begin{center}
    \includegraphics[width=2.25in]{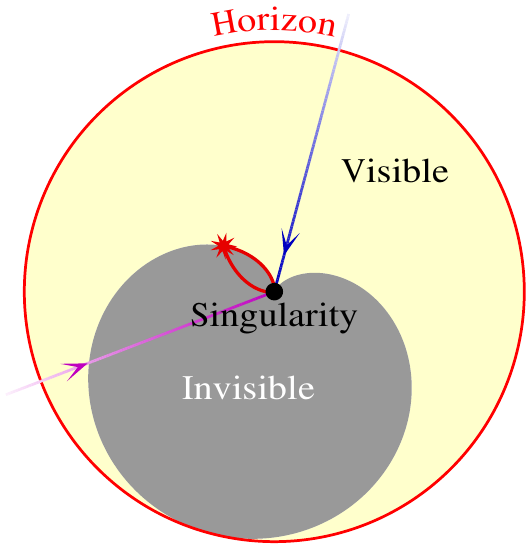}
    \end{center}
    \caption{
The light (yellow) shaded region shows the region visible to an infaller (blue)
who falls radially to the singularity of a Schwarzschild black hole;
the dark (grey) cardioid-shaped
shaded region shows the region that remains invisible to the infaller.
If another infaller (purple) falls along a different radial direction,
the two infallers not only fail to meet at the singularity,
they lose causal contact with each other already some distance
from the singularity.
Since the two infallers fall to two causally disconnected points,
the singularity cannot be a point.
The (red) lines originating at the starred point
show the shortest causal path joining the infallers as they approach
the singular surface.
    }
    \label{schwJoopathvis}
    \end{figure}
}

\newcommand{\schwvizfig}{
    \begin{figure}[tp!]
    \begin{center}
    \leavevmode
    \includegraphics[scale=.087]{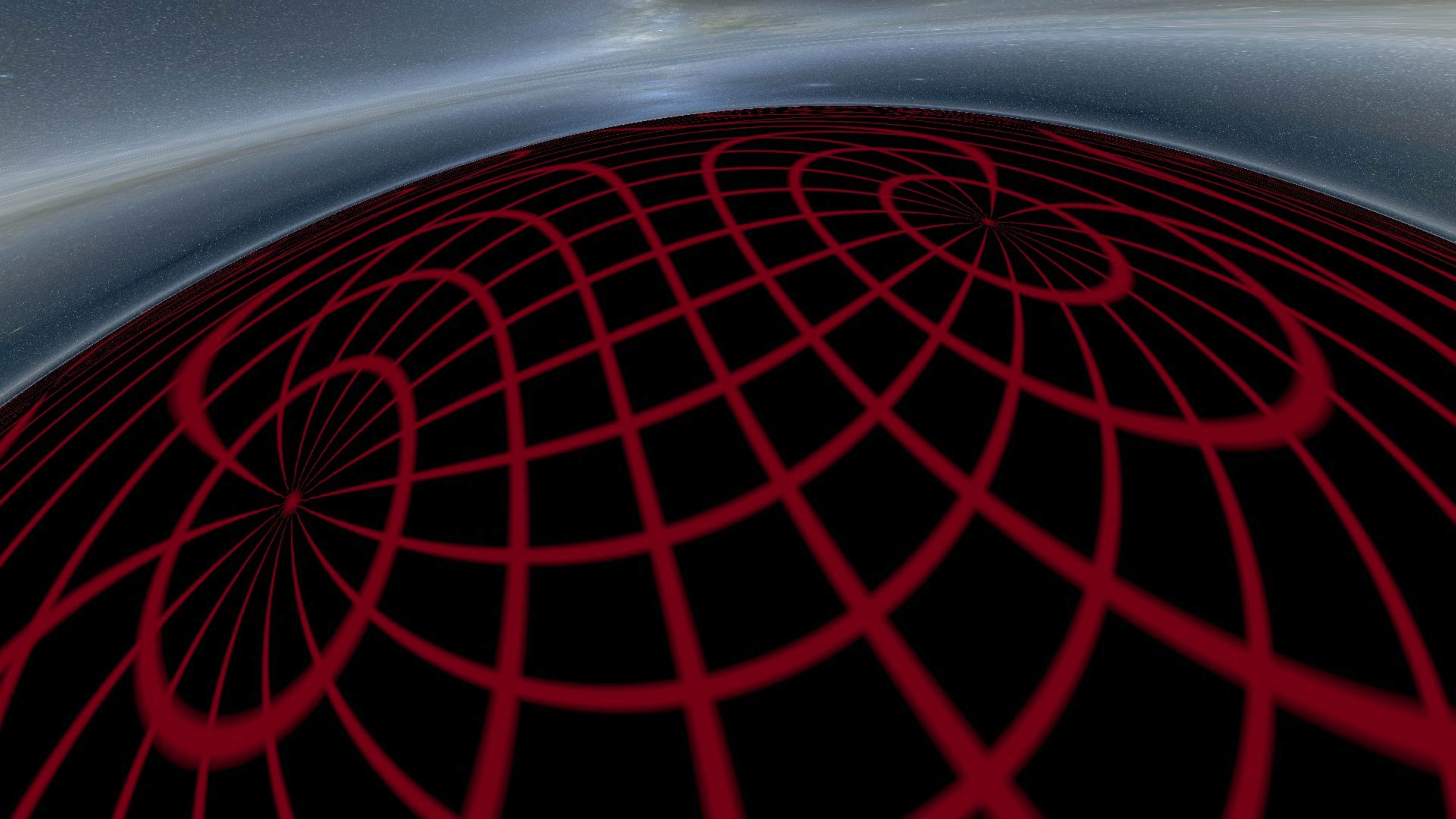}
    \includegraphics[scale=.087]{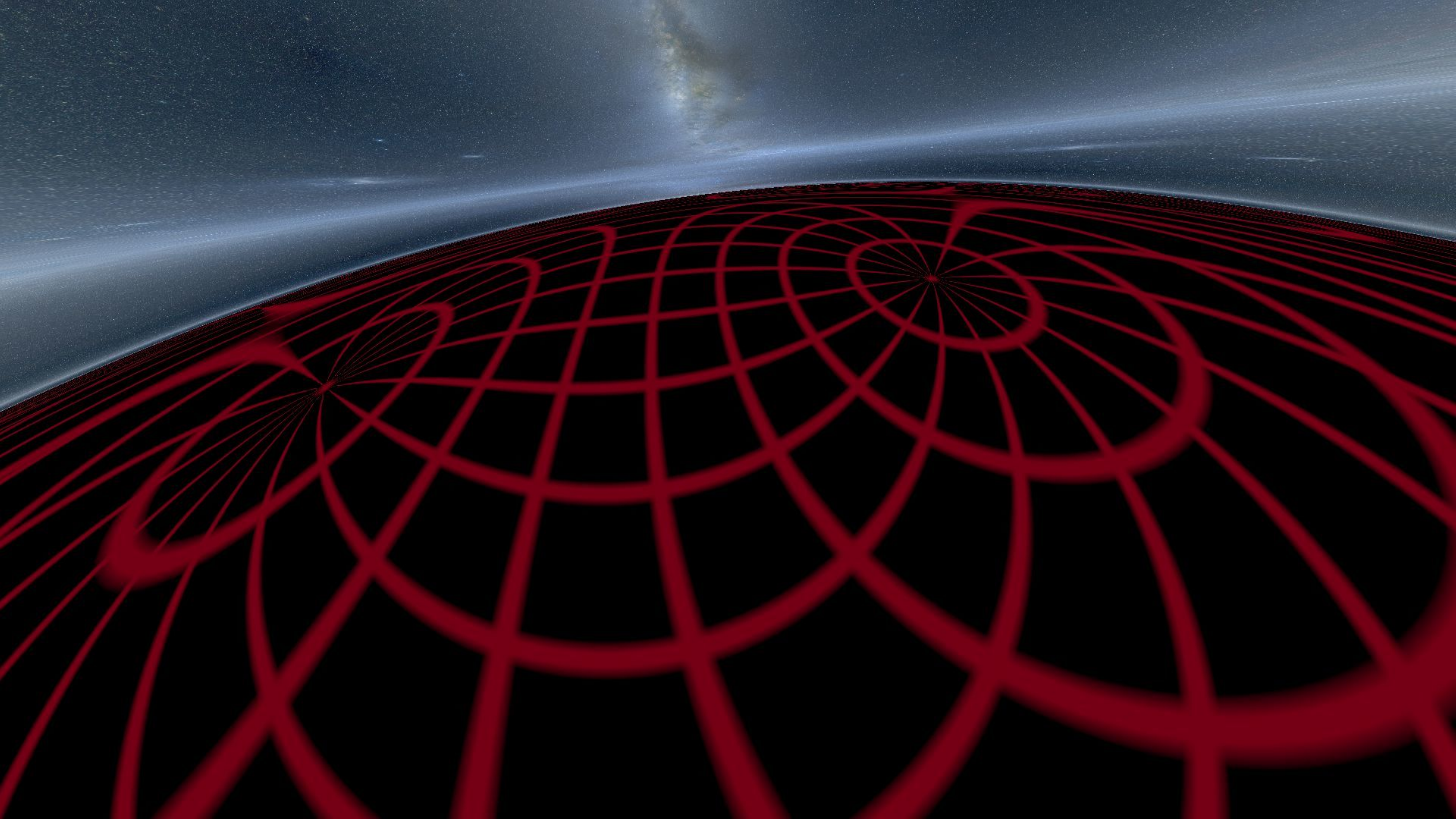}
    \includegraphics[scale=.087]{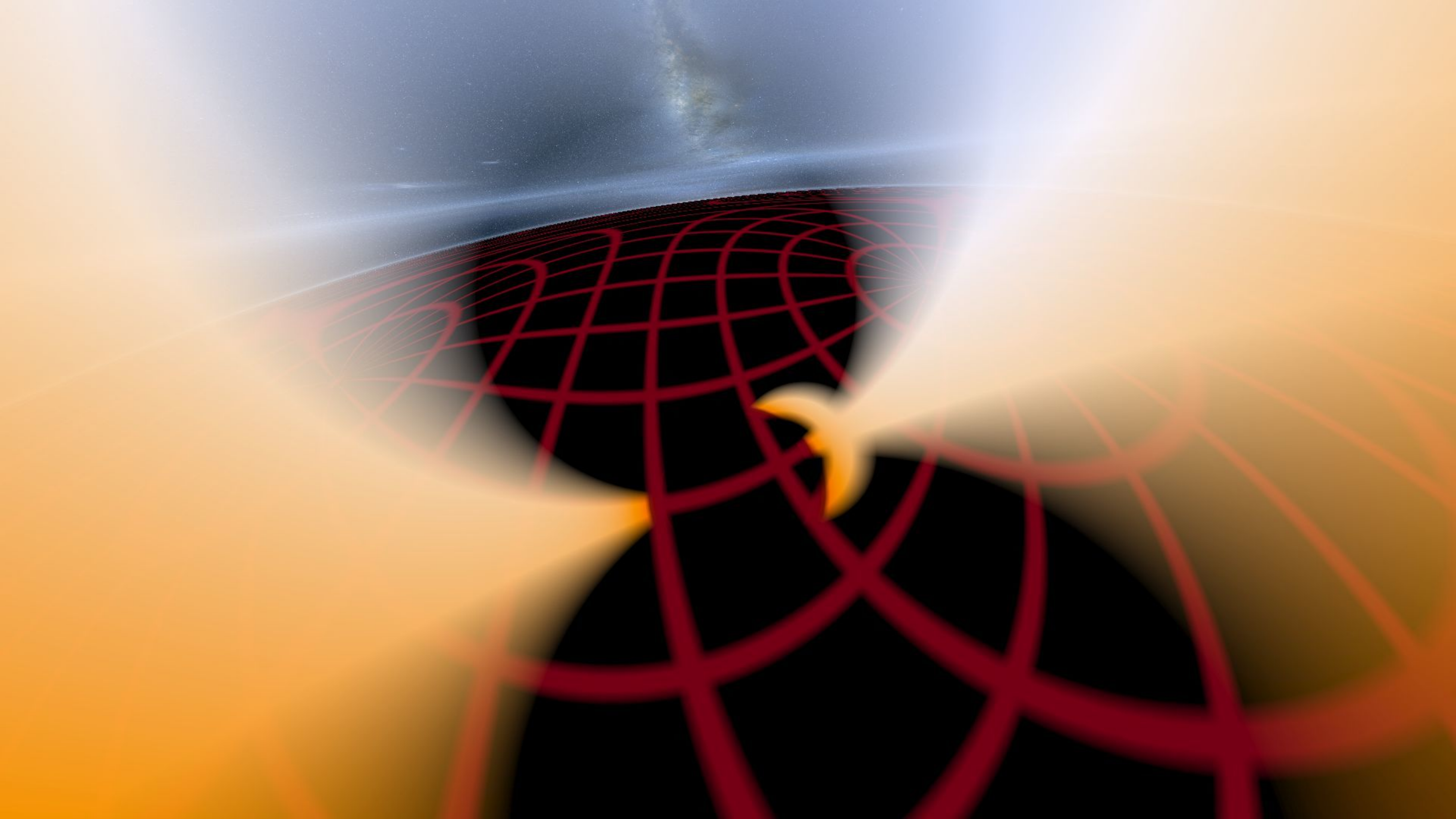}
    \includegraphics[scale=.087]{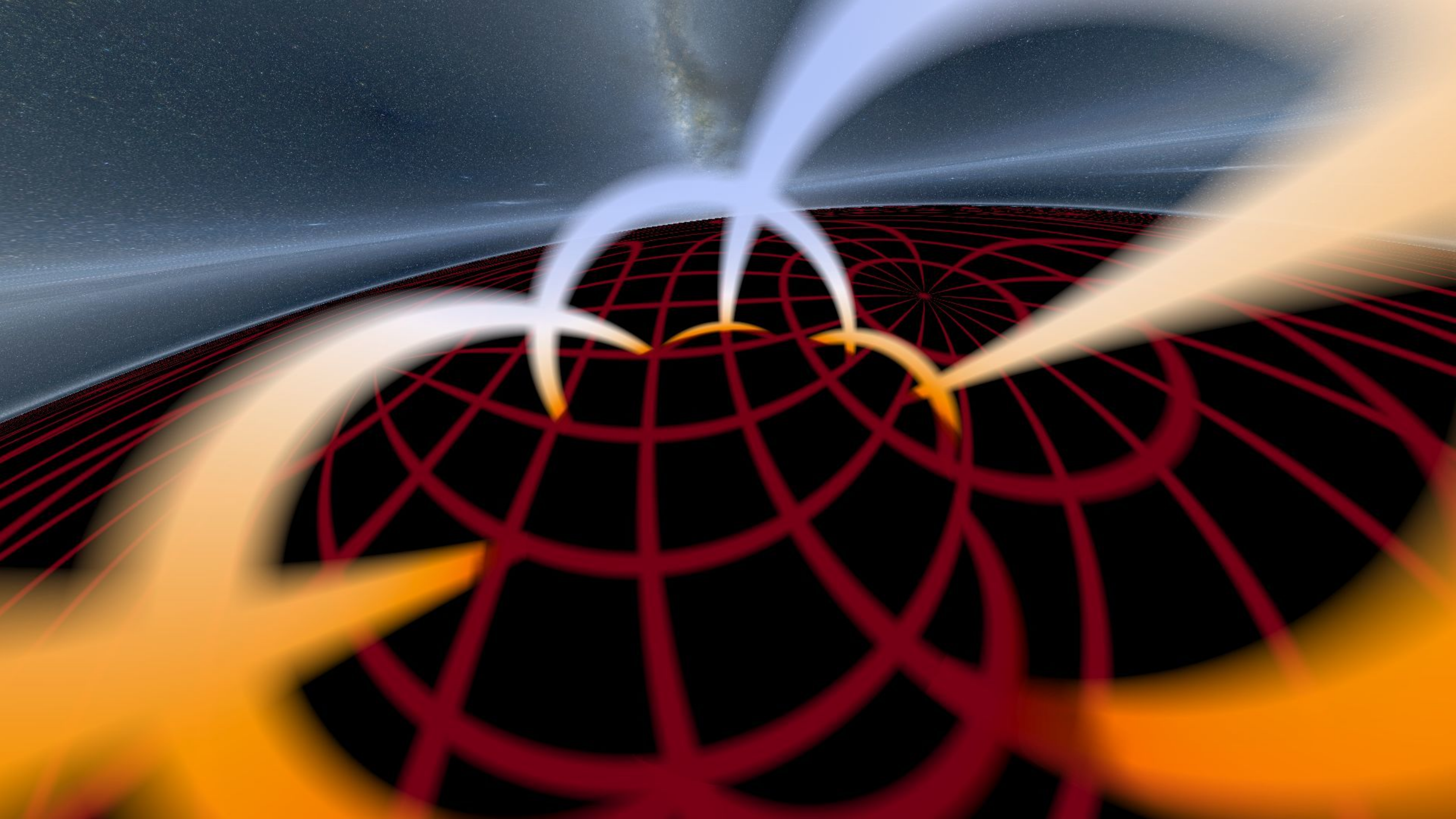}
    \includegraphics[scale=.087]{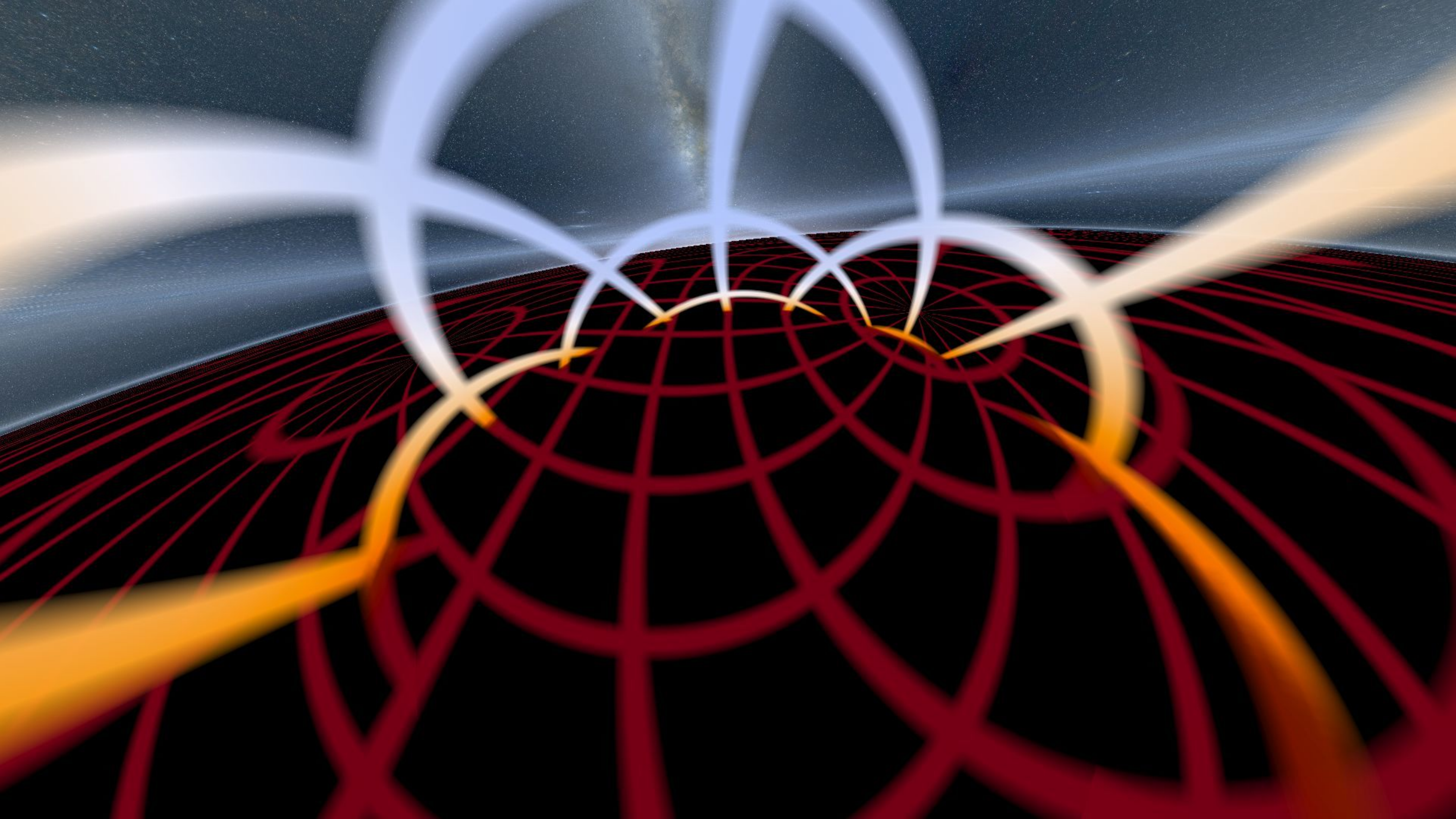}
    \includegraphics[scale=.087]{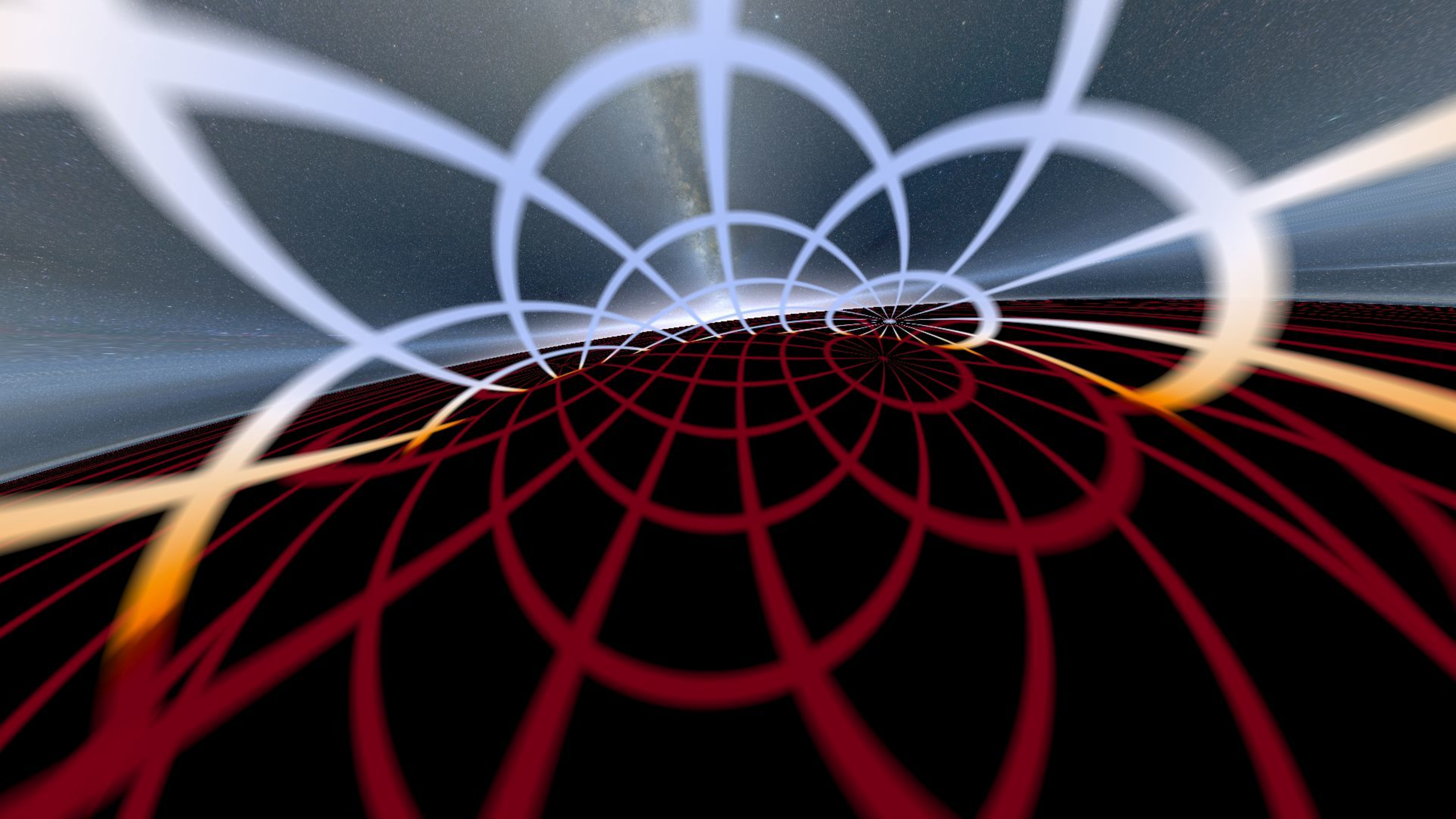}
    \includegraphics[scale=.087]{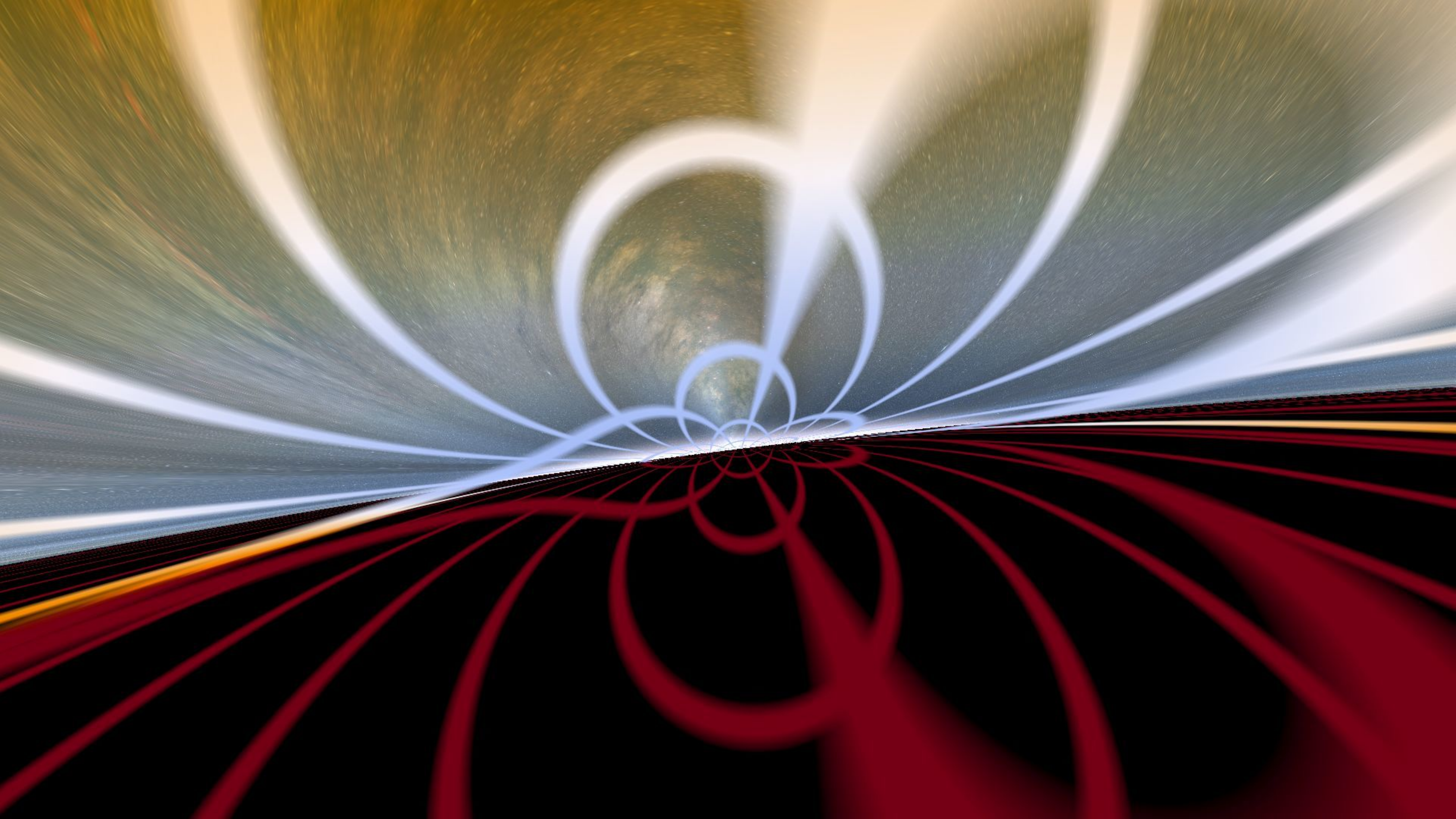}
    \includegraphics[scale=.087]{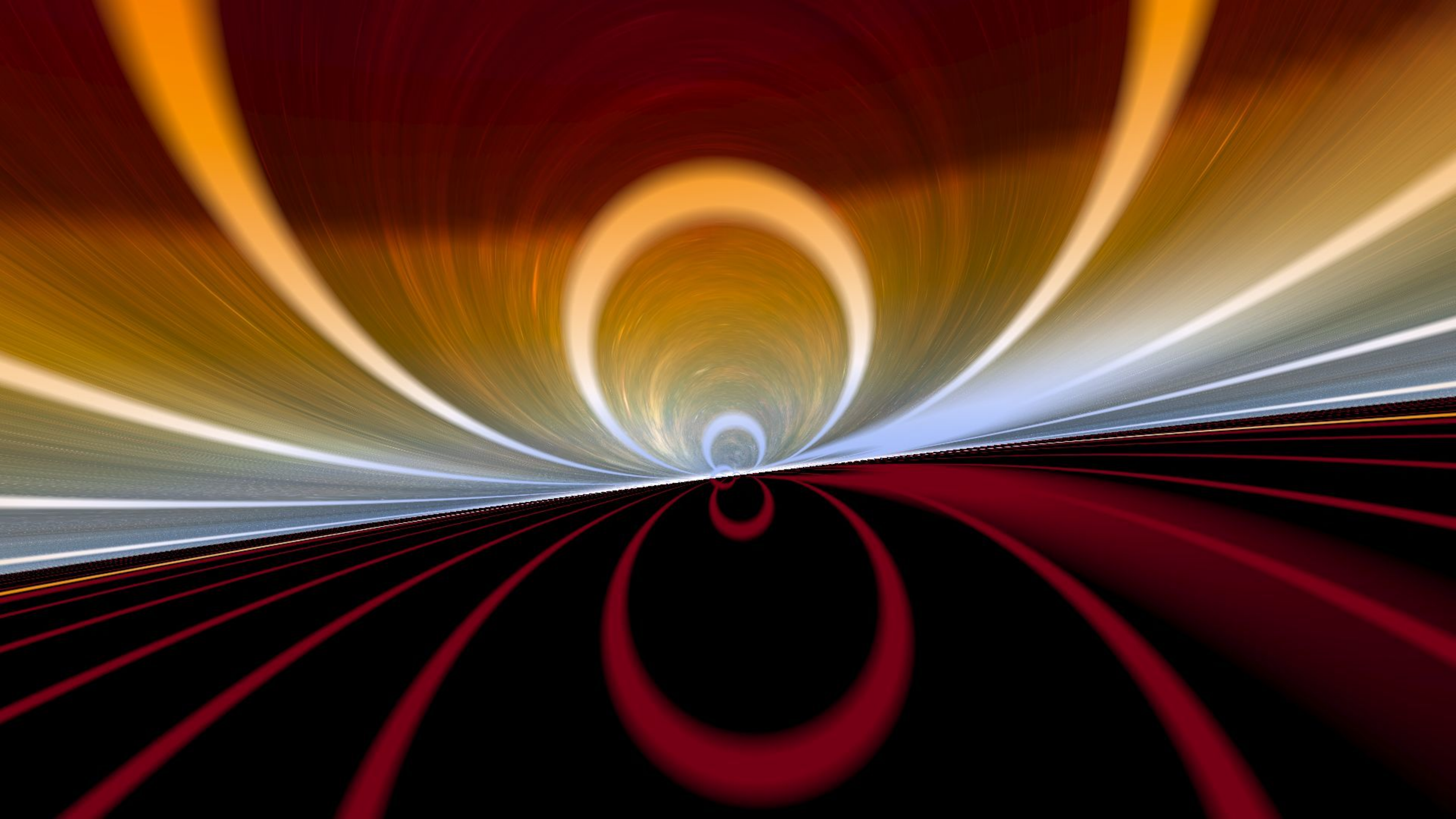}
    \caption[Visualization of falling into a Schwarzschild black hole]{
    \label{schwviz}
Eight frames from a visualization of the view
seen by an observer who free-falls through the horizon of
a Schwarzschild black hole.\cite{Hamilton:2010my}
From left to right and top to bottom,
the observer is at radii
$1.5$,
$1.01$,
$0.99$,
$0.9$,
$0.8$,
$0.5$,
$0.1$,
and
$0.01$
horizon radii.
The illusory horizon is painted with a dark red grid,
while the true horizon is painted with a grid
coloured with an appropriately red- or blue-shifted blackbody colour.
The background is Axel Mellinger's Milky Way.\citep{Mellinger:2009asp}
    }
    \end{center}
    \end{figure}
}

\newcommand{\doranfig}{
    \begin{figure}[b!]
    \begin{center}
    \includegraphics[width=4in]{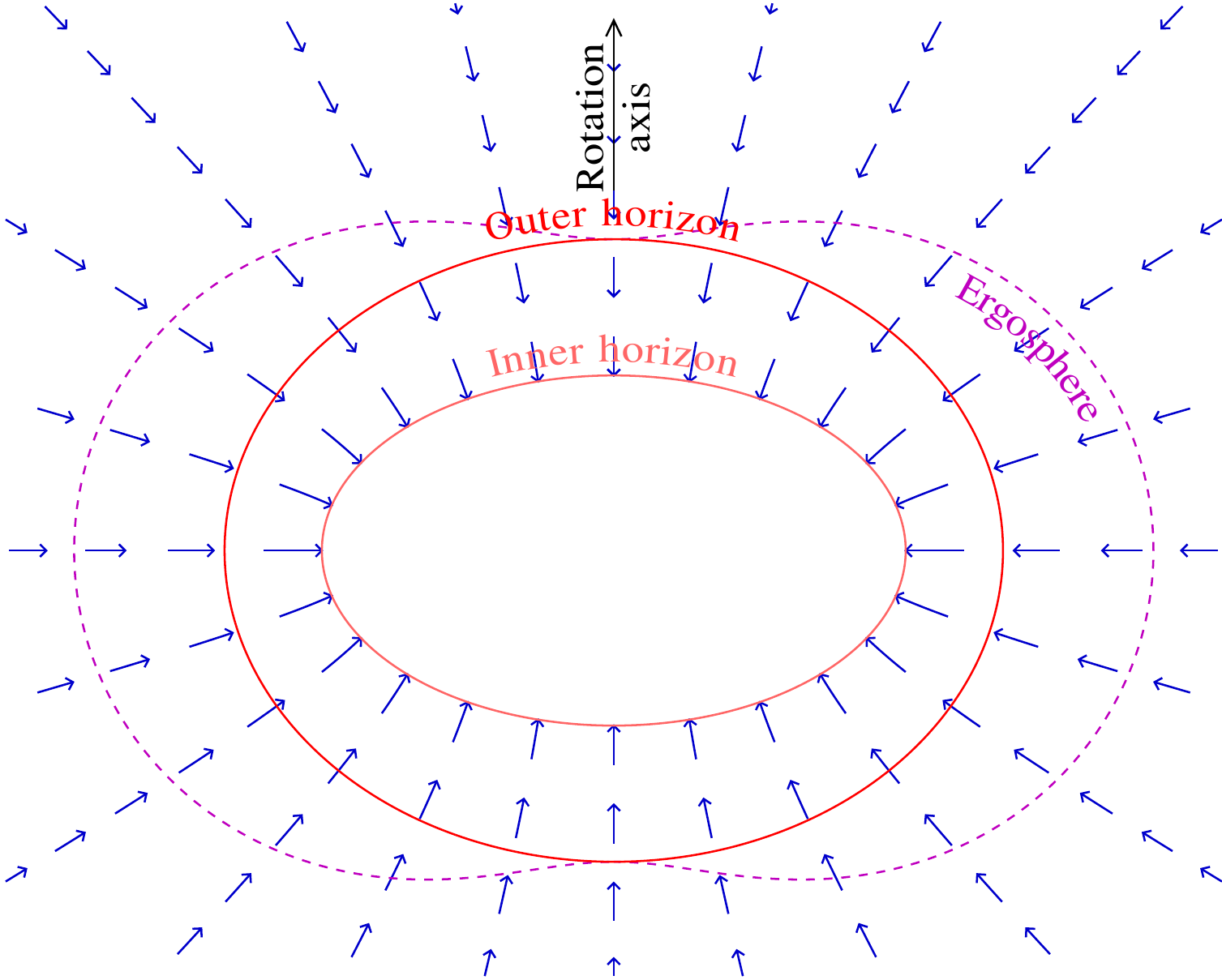}
    \end{center}
    \caption{
Spatial geometry of a Kerr black hole with spin parameter $a = 0.96 M$.
The outer and inner horizons are
confocal spheroids of constant ellipsoidal radius~$r$.
The arrows mark trajectories and velocities
$d r / d \tau$
of infallers who free-fall with zero angular momentum
from zero velocity at infinity.\cite{Doran:1999gb}
The inflow follows lines of constant ellipsoidal latitude~$\theta$,
which form nested hyperboloids orthogonal to and confocal with
nested spheroids of constant ellipsoidal radius.
In a real astronomical black hole,
the Kerr geometry breaks down at the inner horizon
because of the Poisson-Israel mass inflation instability.\cite{Poisson:1990eh}
    }
    \label{doran}
    \end{figure}
}

\newcommand{\penroseschwskyfig}{
    \begin{figure}[t!]
    \begin{center}
    \leavevmode
    \includegraphics[scale=.9]{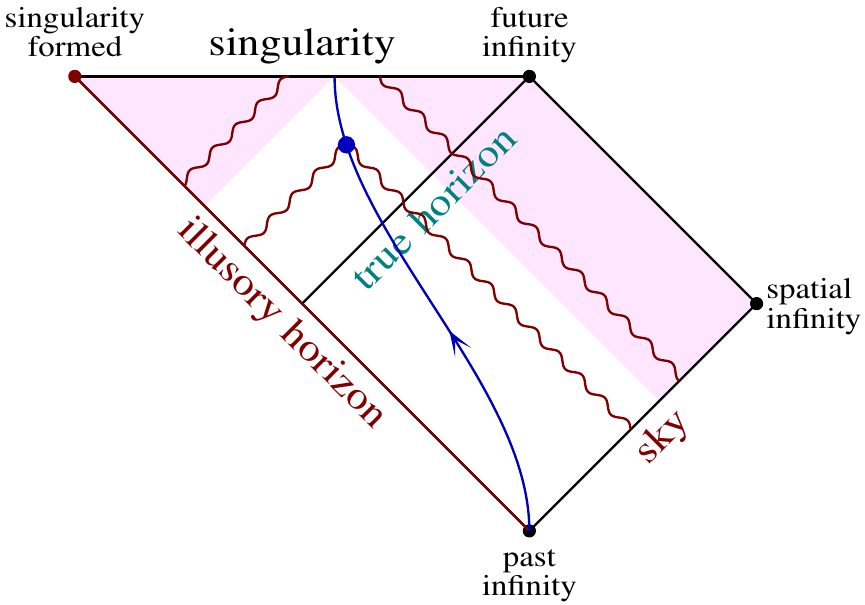}
    \caption[Penrose diagram]{
    \label{penroseschwsky}
Penrose diagram illustrating the trajectory of an observer
who falls to the singularity of a Schwarzschild black hole.
The observer sees Hawking radiation (wiggly lines) from the illusory horizon
below and from the sky above.
The lightly shaded (pink) region surrounds the causal diamond of the observer.
Hawking pair partners lie outside the causal diamond,
ensuring that there is no firewall contradiction.
    }
    \end{center}
    \end{figure}
}

\newcommand{\kappazerofig}{
    \begin{figure}[t!]
    \centering
    \includegraphics[scale=.55]{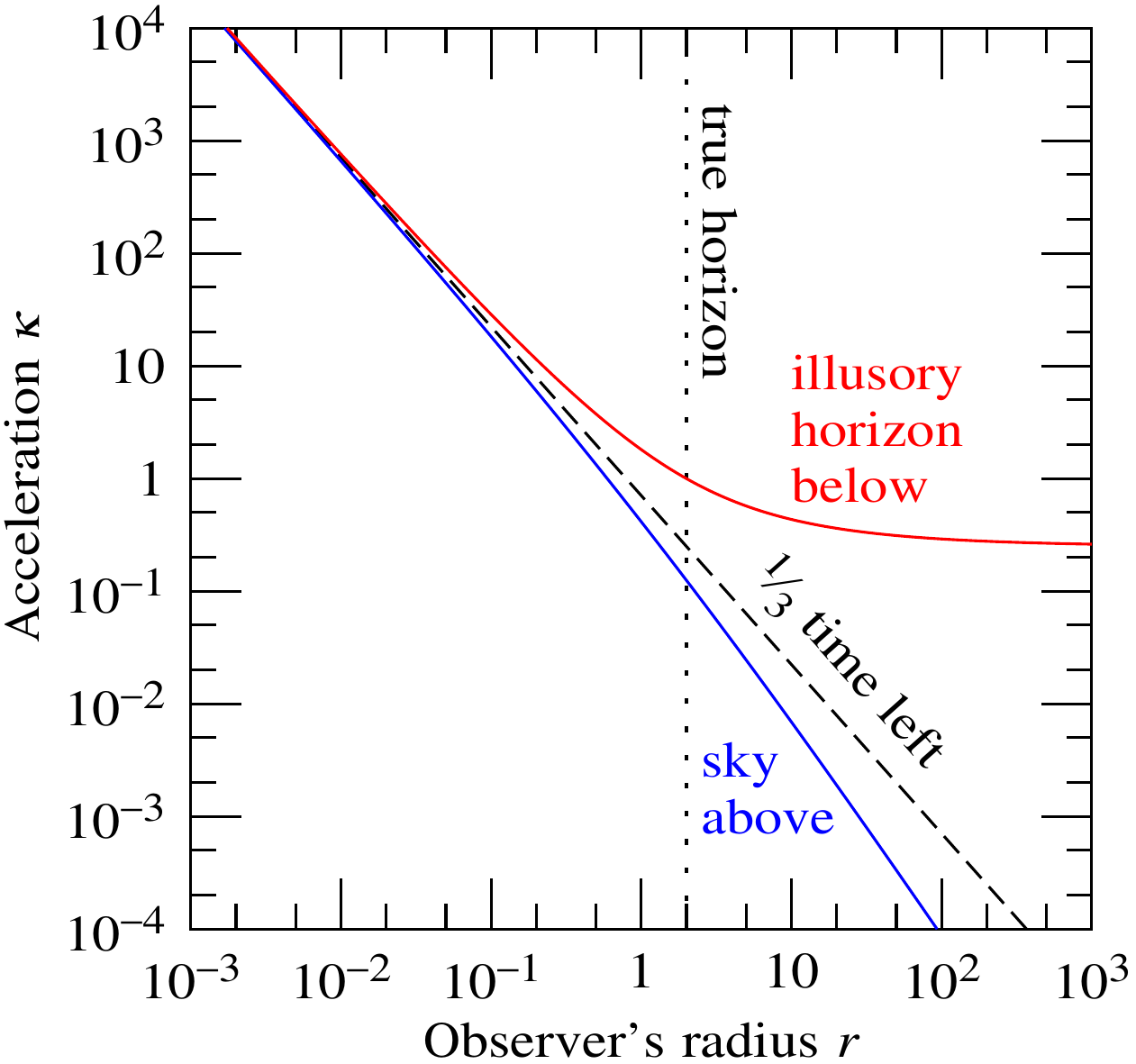}
    \caption[]{
    \label{kappa0}
Acceleration
$\kappa$
on the illusory horizon directly below,
and in the sky directly above,
seen by a radially free-falling infaller at radius $r$.\cite{Hamilton:2016iid}
The units are geometric ($c = G = M = 1$).
Both accelerations asymptote
to one third the reciprocal of the proper time $| \tau |$
left until the infaller hits the singularity,
indicated by the diagonal dashed line.
    }
    \end{figure}
}

\newcommand{\rhofig}{
    \begin{figure}[bt!]
    \centering
    \includegraphics[scale=.55]{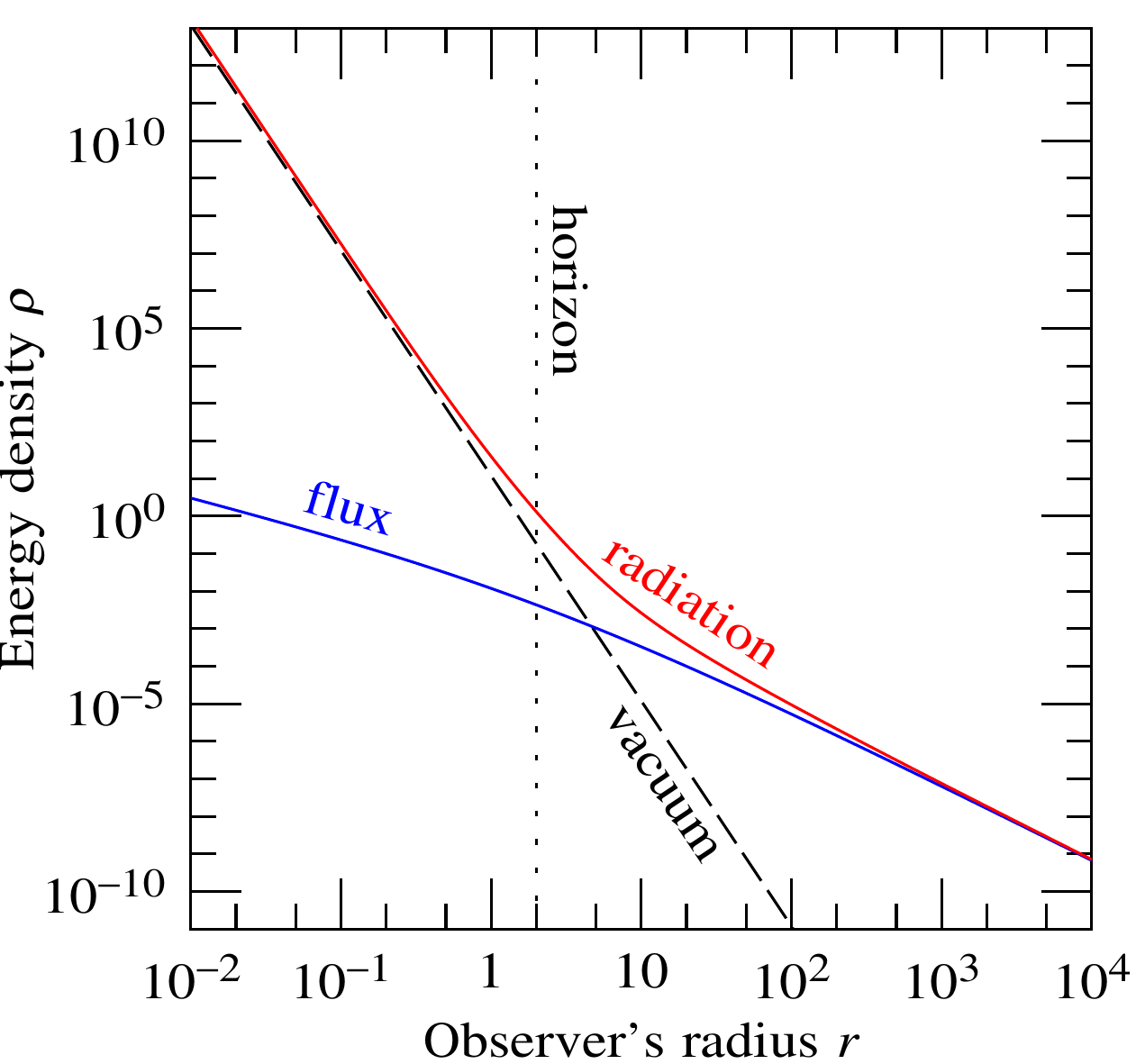}
    \caption[1]{
    \label{rho}
Quantum energy densities $\rho$ of radiation
(solid red line)
and vacuum
(long-dashed black line)
in a Schwarzschild black hole
as a function of radius $r$ in units of $M$.\cite{Hamilton:2016iid}
The radiation density is positive,
while the vacuum energy is negative.
Also shown is
the non-stationary Hawking energy flux
(solid blue line),
which is positive (directed outward).
The vertical dotted line marks the true horizon.
    }
    \end{figure}
}

\newcommand{\penrosemassinflationfig}{
    \begin{figure}[t!]
    \begin{center}
    \leavevmode
    \includegraphics[width=3.5in]{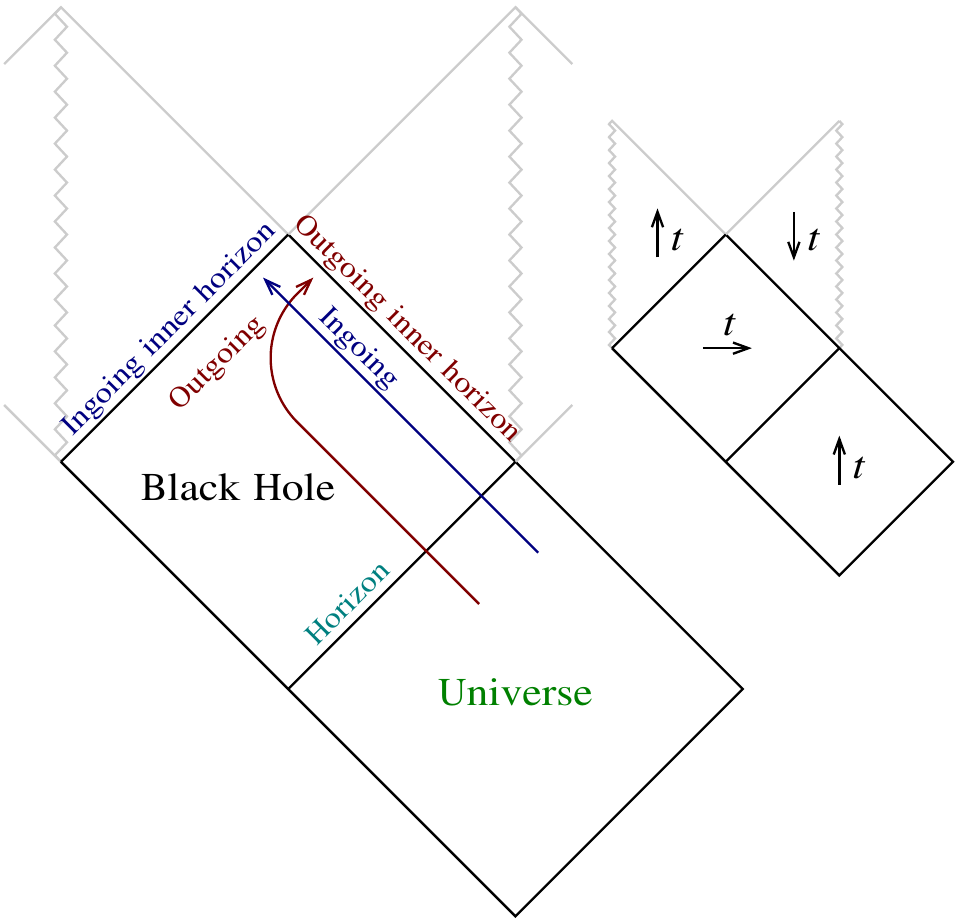}
    \caption{
Penrose diagram illustrating why Kerr black holes
are subject to the Poisson-Israel\cite{Poisson:1990eh}
mass inflation instability.
Ingoing (retrograde) and outgoing (prograde) streams just outside
the inner horizon must pass through separate ingoing and outgoing inner horizons
into causally separated pieces of spacetime where the
timelike time coordinate $t$ goes in opposite directions.
To accomplish this, the ingoing and outgoing streams must
exceed the speed of light through each other,
which physically they cannot do.
In a real black hole,
the energy-momentum of
hyper-relativistically counter-streaming ingoing and outgoing streams
just above the inner horizon back-reacts on the geometry,
leading to inflation and collapse,
and cutting off the analytic continuations
through the inner horizons of the Kerr geometry.
The inset shows the direction of coordinate time $t$
in the various regions.
Proper time of course always increases upward
in a Penrose diagram.
    }
    \label{penrosemassinflation}
    \end{center}
    \end{figure}
}

\newcommand{\rninnerhorizonfig}{
    \begin{figure}[t!]
    \begin{center}
    \leavevmode
    \includegraphics[width=4.8in]{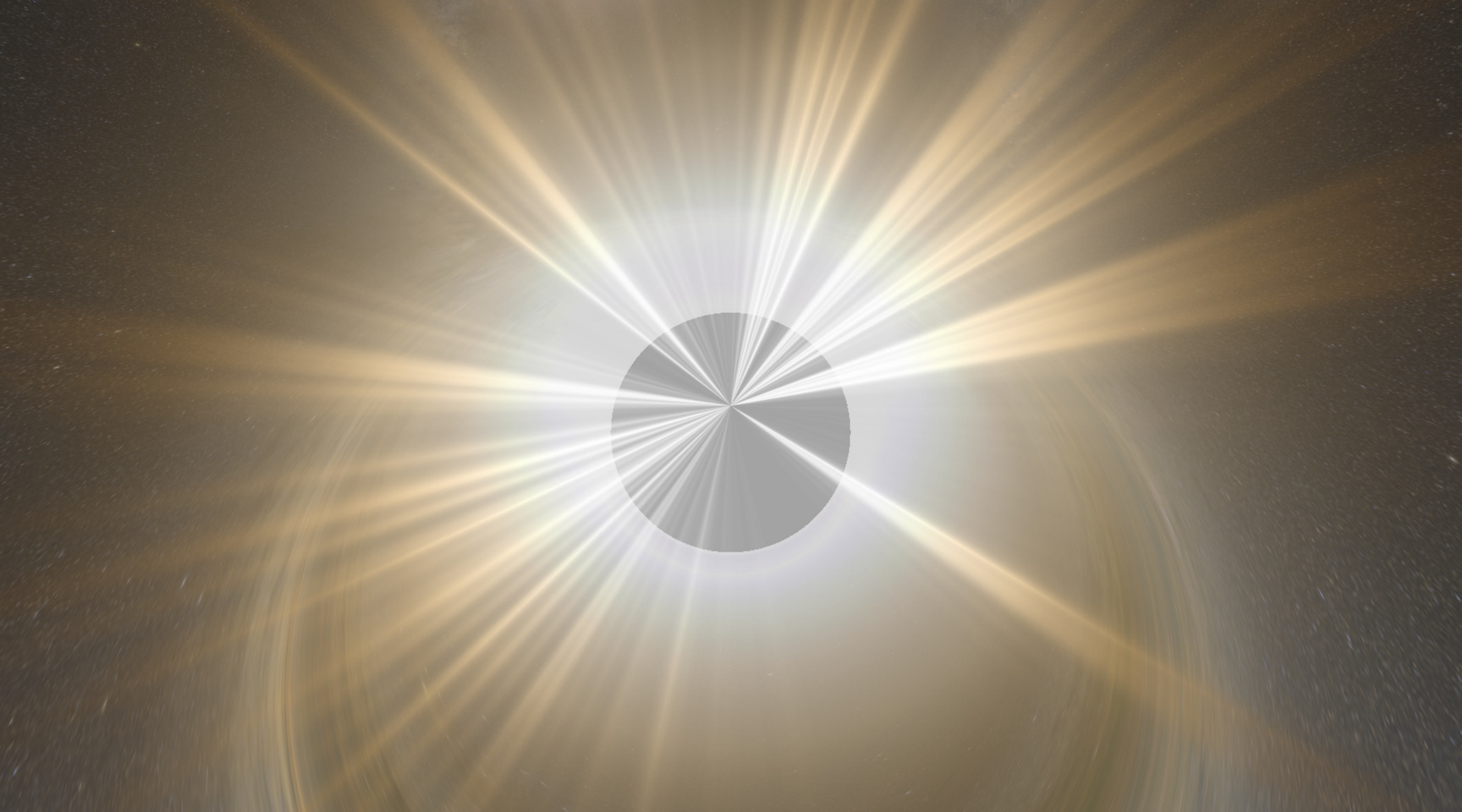}
    \end{center}
    \caption[Orbits asymptoting to unstable circular orbits]{
    \label{rninnerhorizon}
An ingoing observer at the inner horizon of
a Reissner-Nordstr\"om (spherical, charged) black hole sees
outgoing light and matter focused into a divergingly blueshifted bright point.
See \url{https://jila.colorado.edu/~ajsh/insidebh/rn.html}.
The background is Axel Mellinger's Milky Way.\citep{Mellinger:2009asp}
    }
    \end{figure}
}

\newcommand{\inflationaryhorizonfig}{
    \begin{figure}[t!]
    \begin{center}
    \leavevmode
    \includegraphics[width=2.8in]{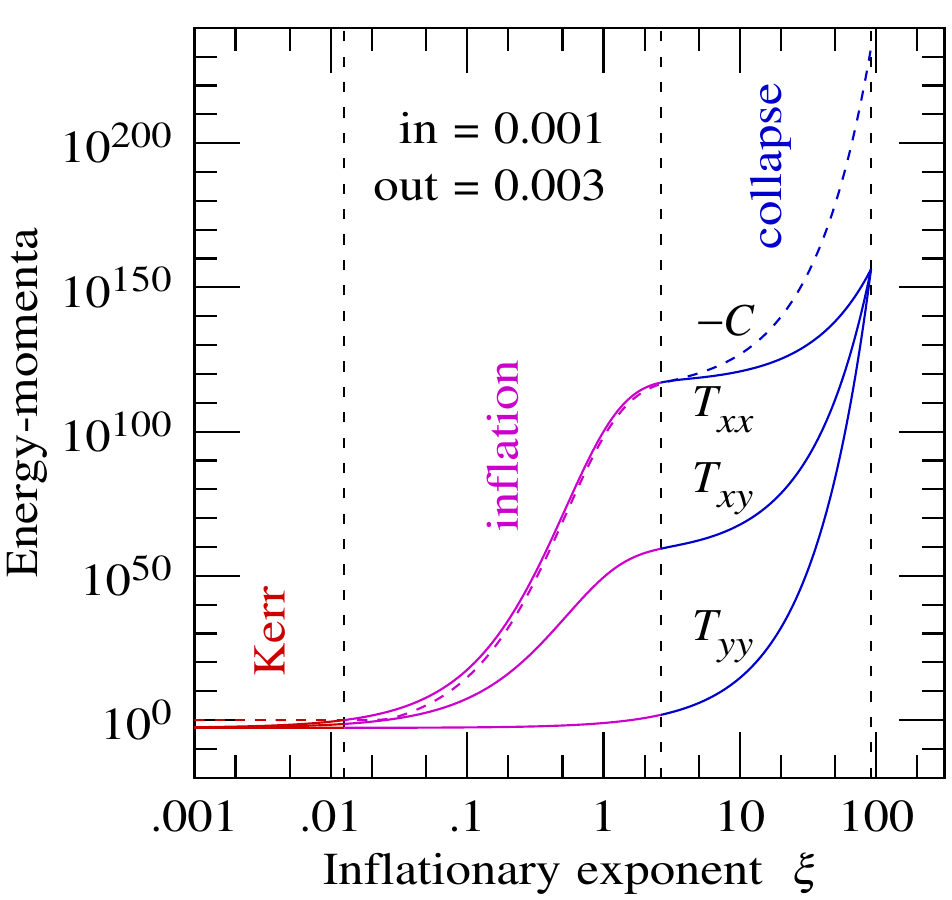}
    \caption[1]{
    \label{inflationaryhorizon}
Tetrad-frame
radial, radial-angular, and angular collisionless energy-momenta
$T_{xx}$,
$T_{xy}$,
and
$T_{yy}$
in a conformally separable, accreting, rotating black hole,
from the initial Kerr regime,
through inflation to collapse.\cite{Hamilton:2010a}
The energy-momenta grow exponentially huge
despite their small initial values.
The dashed line is minus the polar (real) spin-$0$
component of the Weyl curvature
$- C$.
The axial (imaginary) spin-$0$ Weyl component
is comparable to $T_{yy}$.
    }
    \end{center}
    \end{figure}
}

\newcommand{\kninfafig}{
    \begin{figure}[t!]
    \begin{center}
    \includegraphics[width=3in]{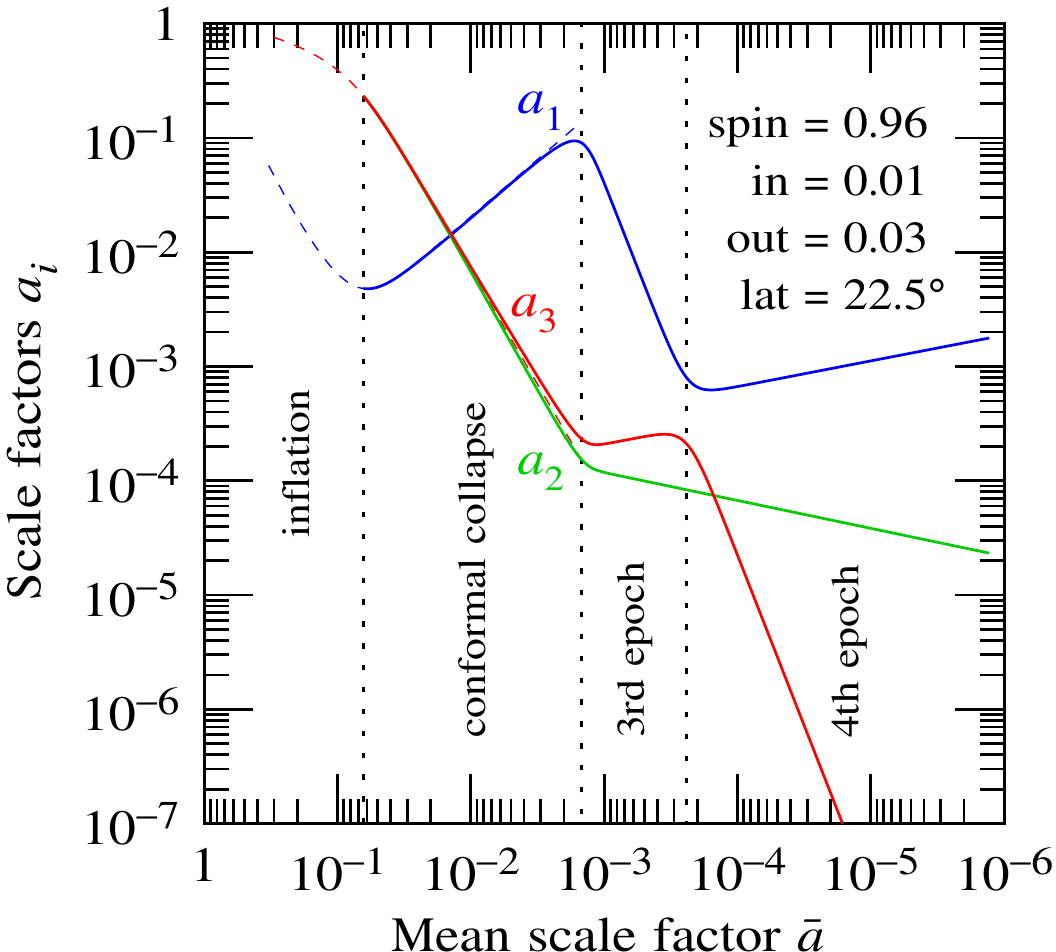}
    \end{center}
    \caption{
Numerical calculation of an accreting, rotating black hole
from inflation to oscillatory BKL collapse.
The numerical integration (solid lines)
took the conformally separable solution (dashed lines) as initial condition.
There are 4 power-law Kasner epochs before the numerical integration fails.
The agreement between the solid and dashed lines during conformal collapse
supports the correctness of the numerical and conformally separable solutions.
    }
    \label{kninfa}
    \end{figure}
}

\newcommand{\kninfqfig}{
    \begin{figure}[t!]
    \begin{center}
    \includegraphics[width=3in]{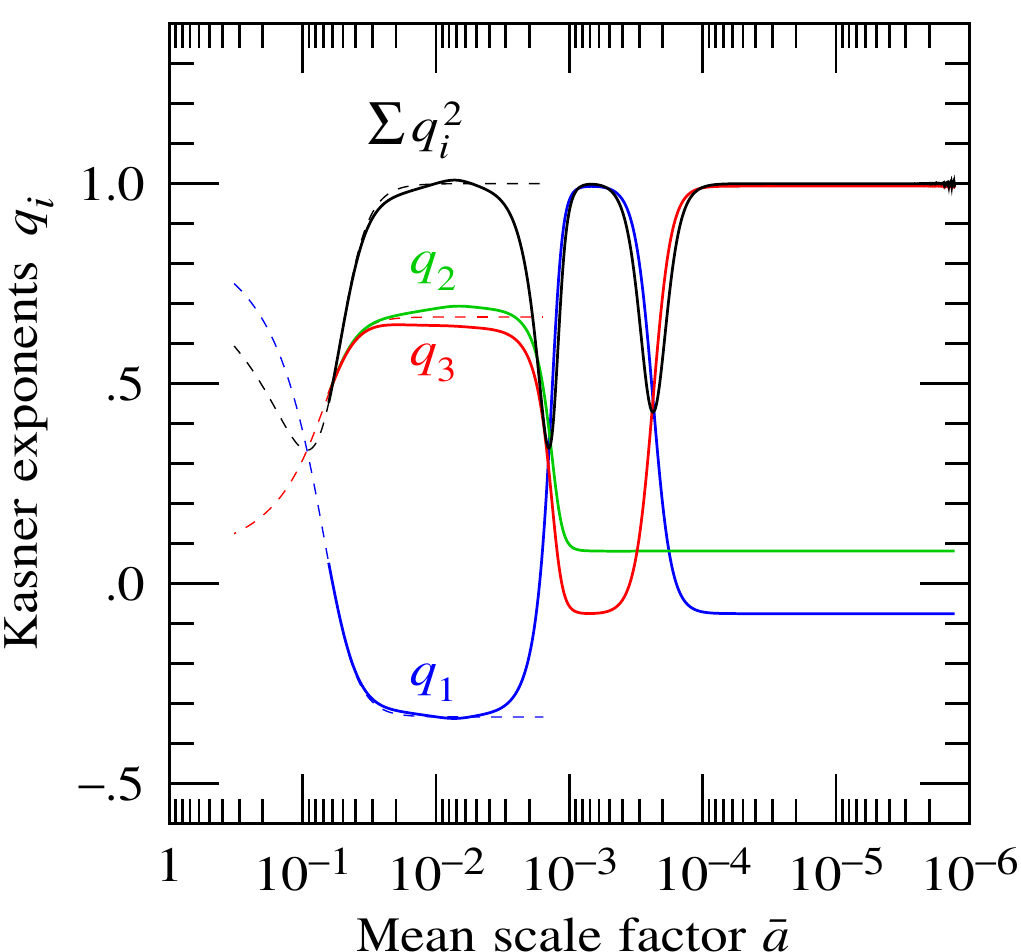}
    \end{center}
    \caption{
The Kasner exponents $q_i$
corresponding to the evolution shown in Figure~\ref{kninfa}.
The Kasner exponents of the second epoch coincide with those of Schwarzschild.
    }
    \label{kninfq}
    \end{figure}
}

\title{Inside astronomically realistic black holes}
\author{Andrew J. S. Hamilton}

\address{JILA, U.\ Colorado Boulder, \\
Boulder, CO 80309, USA \\
$^*$E-mail: Andrew.Hamilton@colorado.edu}

\begin{abstract}
The singularity of a spherical (Schwarzschild) black hole is a surface,
not a point.
A freely-falling, non-rotating observer sees Hawking radiation
with energy density diverging with radius as
$\rho \propto r^{-6}$ near the Schwarzschild singular surface.
Spacetime inside a rotating (Kerr) black hole terminates at the inner
horizon because of the Poisson-Israel mass inflation instability.
If the black hole is accreting, as all realistic black holes do,
then generically inflation gives way to
Belinski-Khalatnikov-Lifshitz
oscillatory collapse to a strong, spacelike singular surface.
\end{abstract}


\bodymatter


\section{Introduction}

What really happens inside black holes?
The question is of perennial fascination to students and the public.
Yet the literature on the subject is surprisingly sparse.
That is to say,
there is a large literature on singularities and quantum gravity,
but little on the more prosaic topic of what happens on the way
to a singularity.
In fact it's a wilder ride than you might imagine.

What would {\em you\/} respond to a student or member of the public
who asks what happens inside black holes?
Here is a sample of possible answers:
\begin{enumerate}
\item
Nothing can escape a black hole, so what happens inside is unknowable.
\item
\label{point}
``All that makes up a black hole
is crushed together at a single miniscule point at the black hole's center.''
\item
You are spaghettified.
\item
You are burned up by a firewall at the horizon.
\item
You are burned up by Hawking radiation as you approach the singularity.
\item
There is a weak null singularity on the Cauchy horizon.
\item
There is BKL oscillatory collapse to a spacelike singularity.
\item
There is a naked singularity.
\end{enumerate}
Here are my opinionated answers to the above questions:
(1) False---see~\S\ref{future}.
(2) False---see~\S\ref{notapoint}.
(3) True.
Interesting fact:
you are spaghettified a tenth of a second before you hit the singularity
(tidal force and time both depend in the same way $GM/r^3$
on mass $M$ and radius $r$).
(4) False---see~\S\ref{horizonnotabarrier}.
(5) True---see~\S\ref{hawkingschwarzschild}.
(6) False, at least for astronomically realistic black holes---see~\S\ref{inflation}.
(7) True---see~\S\ref{bkl}.
(8) I suspect that classical naked singularities do not survive quantum back reaction.

\section{The Singularity is Not a Point}
\label{notapoint}

It is widely repeated in popular and indeed expert writing
that the singularity at the centre of a (spherical) black hole is a point.
The quote in item~(\ref{point}) of the Introduction is from
Brian Greene's ``The Fabric of the Cosmos'', page~337.\cite{Greene:2004}
It's not true.

Of course, a mathematical relativist would say, correctly,
that the central point is not part of the spacetime,
and should be excised.
Rigorously, in general relativity one should make statements about
behaviour asymptotically close to, but not at, the singularity.
But that is not the point I am making.
The singularity is a surface, not a point.
Let me prove it to you.

\schwJoopathvisfig

Figure~\ref{schwJoopathvis}
shows two infallers, blue and purple, who free-fall radially
to the singularity of a Schwarzschild black hole.
The defining feature of black holes is that space falls faster than light
inside their horizons\cite{Hamilton:2004au}
(the rule is that nothing can move {\em through\/} space faster than light;
spacetime itself can do whatever general relativity prescribes).
So inside the horizon, light must move inward,
regardless of the direction in which it may be pointed.
There are lightrays that have the maximum possible sideways motion
(technically, these lightrays have infinite angular momentum per unit energy).
The cardioid-shaped\footnote{The region is in fact a mathematical cardioid.}
grey region in Figure~\ref{schwJoopathvis} is bounded by
maximally-sideways-moving lightrays that just hit the blue infaller
as the latter hits the singularity.
Anything in the grey region is invisible to the blue infaller:
a signal emitted in the grey region would have to move through space
faster than light to reach the infaller at (just outside)
the singularity.

Therefore the last that the blue infaller sees of the purple infaller
is at the boundary of the grey region, already well away from the singularity.
Far from encountering each other at the singularity,
the blue and purple infallers lose causal contact with each other
already some distance from the singularity.
Since the two infallers fall to causally disconnected places,
the singularity cannot be a point:
it must be a surface (as will be elucidated in a moment).

You might object that the spatial separation between the two infallers
goes to zero at the singularity,
as is evident from the Schwarzschild metric.
The counter-intuitive thing here is that,
contrary to familiar experience,
being spatially close does not imply being causally close.
The two red lines originating from the starred point in
Figure~\ref{schwJoopathvis}
show the shortest causal path joining the two infallers
as they approach the singularity.
A pair of lightrays emitted from the starred point would just reach the
infallers as they hit the singularity.

If the Schwarzschild singularity is a surface,
what is the nature of that surface?
Classically, the boundary of a 4-dimensional spacetime is 3-dimensional.
Since it is the time dimension that gives up the ghost at the Schwarzschild
singularity, classically the boundary is spacelike.
However,
quantum mechanically, if unitarity holds,
then the singularity at different times
(as seen by observers who fall in at different times)
must be unitarily related to each other.
Ordinarily, unitary evolution is what happens along a time dimension.
Therefore an effective time dimension must emerge,
and the singular surface should be thought of as a 2-dimensional
surface that evolves unitarily in time.
The notion that the singularity is a 2-dimensional surface accords
with what Hawking taught us, that a black hole is a thermodynamic object
with number of states proportional to the 2-dimensional area of its horizon.

\section{The Horizon is Not a Barrier}
\label{horizonnotabarrier}

\schwvizfig

An enduring feature of (mis)understanding of black holes is ongoing confusion
over the nature of horizons.
The classical confusion was cleared up by
David Finkelstein in 1958,\cite{Finkelstein:1958zz}
who showed that light falls freely through the horizon,
and does not get stuck at the horizon as a simplistic interpretation
of the Schwarzschild metric had suggested.

In modern times the confusion has reappeared in a quantum context.
Hawking radiation is emitted from the horizon,
and the entropy of a black hole equals a quarter of its horizon area.
What happens to an observer who falls to the horizon?
Do they see Hawking radiation there?
Do they encounter the states of the black hole there?

The clearest example of quantum confusion over the horizon is
the ``firewall paradox'' posited by
Almheiri, Marolf, Polchinski, and Sully in 2012.\cite{Almheiri:2012rt}
More on this at the end of this section.

Consider the following thought experiment.
You free-fall feet-first into a black hole.
What do you see as you look down at your feet when your feet pass through
the horizon?
Obviously, since you are in a free-fall frame,
you continue to see your feet below you.
The light emitted by your feet upwards at the horizon remains at the horizon,
and your eyes catch up with the light from your feet
as your eyes fall through the horizon,
but you do not catch up with your feet themselves.
Continuing to free-fall inside the horizon,
you continue to see your feet below you.
Inside the horizon,
where space is falling faster than light,
you see an image of your feet below you as they used to be above your head.

The moral of this story is that when you fall through the horizon,
you do not catch up with the source of any light
that may have been emitted upwards at the horizon.
For example,
when you watch a black hole from outside the horizon,
what you are looking at is not actually its horizon,
but rather the dimming, redshifting surface of the star or whatever else
collapsed to, or fell into, the black hole long ago.
When you free-fall through the horizon, you do not catch up with
the star or other stuff that collapsed long ago.
It's already gone.
Its image remains ahead of you, still dimming and redshifting away.

Ray-tracing confirms these arguments.
Figure~\ref{schwviz}
shows eight frames from a visualization\cite{Hamilton:2010my} of the view
seen by an observer who free-falls through the horizon of
a Schwarzschild black hole.
The red grid is nominally the past horizon.
A real black hole formed from stellar collapse does not have a past horizon.
Rather, the past horizon is replaced by the dimming, redshifting surface
of the star that collapsed long ago.
Gavin Polhemus and I\cite{Hamilton:2010my} started calling this surface
the ``illusory horizon''
to distinguish it from the true horizon
that you actually fall through.

An observer who free-falls through the true horizon
does not encounter a concentration of classical radiation emitted there
by the star that collapsed long ago,
and no more so do they encounter Hawking radiation there.
The dimming, redshifting surface of the star
still appears ahead of the observer,
and so also does the Hawking radiation.
The illusory horizon has all the properties that a source of Hawking
radiation should have:
it is the boundary that divides what an observer can see from what
they cannot see;
and from the perspective of the observer,
objects infalling through the illusory horizon
continue to redshift exponentially,
albeit at an exponential rate that varies with time.

Where do the arguments of the firewall paradox go wrong?
The fundamental problem is a confusion between the true and illusory horizons.
Any argument involving regions outside and inside ``the'' horizon,
for example quantum entanglement between outside and inside, must
take into account that the notion of outside and inside is observer-dependent.
It is the illusory horizon, not the true horizon,
that divides what an observer can see from what they cannot.
The ``near-horizon zone'' that so vexes firewallers is observer-dependent,
just as it is in cosmology.

Figure~\ref{schwviz}
illustrates that a free-faller has the impression of reaching the illusory
horizon as they hit the singularity.
At the singularity,
the illusory horizon has the appearance of a flat, 2-dimensional surface.
This is where an infaller finally encounters the quantum states of
the black hole.
This is where the firewall resides.

\section{Hawking Radiation inside a Schwarzschild Black Hole}
\label{hawkingschwarzschild}

\penroseschwskyfig

Hawking radiation
is widely considered to be a profound clue to quantum gravity.
There are thousands of papers on Hawking radiation as seen by observers
outside the horizon of a black hole,
but practically nothing from the perspective of an observer who
falls inside the horizon.
This seems mysterious: if Hawking radiation is such a profound clue,
would one not learn more by exploring it near the singularity?

Part of the difficulty is that the problem of Hawking radiation inside a
black hole is mathematically challenging.
Following Hawking's original 1974 paper,\cite{Hawking:1974} there was
a spate of work on quantum field theory in curved spacetime.\cite{Birrell:1982}
There are two separate but related questions:
(1) What does an observer see?
(2) What is the quantum modification to the energy-momentum tensor?
The first question is complicated by the fact that an accelerating observer
sees quantum radiation---Unruh\cite{Unruh:1976db} radiation---that
a free-faller does not.
As regards the second question,
a rigorous approach to calculating the quantum energy-momentum
in a given curved spacetime
involves summing over an infinite series of divergent terms.
The divergent part of each term must be isolated and subtracted
before the series can be summed.
Unfortunately the problem proved intractable, even in the simple case
of the Schwarzschild geometry.

However, some exact results are calculable.
The so-called trace anomaly $\langle T_\mu^\mu \rangle$, or conformal anomaly,
which is the trace of the expectation value of the quantum energy-momentum,
equals a certain combination of squares of Riemann curvature components.
For a Schwarzschild black hole of mass $M$,
the trace anomaly is\cite{Duff:1993wm}
(in Planck units $c = G = \hbar = 1$)
\begin{equation}
\label{traceanomaly}
  \langle T_\mu^\mu \rangle
  =
  {q_{\rm eff} \over 60\pi^2}
  {M^2 \over r^6}
  \ ,
\end{equation}
where $q_{\rm eff}$ is an effective number of particle species.
The trace anomaly~(\ref{traceanomaly}) diverges as $r^{-6}$
near the singular surface.
Further calculable constraints emerge from the requirement that
the expectation value $\langle T_{\mu\nu} \rangle$
of the quantum energy-momentum be covariantly conserved,
an approach pioneered by Christensen et al.\cite{Christensen:1977}

There is a second reason why there has been so little progress on
Hawking radiation inside the horizon,
namely confusion between the illusory and true horizons.
As remarked in \S\ref{horizonnotabarrier} above,
it is the illusory horizon that is the source of Hawking radiation,
for both inside and outside observers.
The illusory horizon is the surface where
an observer see objects redshifting exponentially.

Although a full quantum calculation is still intractable,
it is relatively straightforward to go to the classical high-frequency
limit of geometric optics,
and ray-trace from the illusory horizon to the observer.
A quantity of central interest is the acceleration $\kappa$,
the rate of redshifting at the illusory horizon.
If the acceleration $\kappa$ were constant,
then the standard Hawking calculation shows that the
emission would be thermal with temperature $T$ equal to the acceleration
divided by $2\pi$,
\begin{equation}
  T = {\kappa \over 2\pi}
  \ .
\end{equation}

\kappazerofig

Figure~\ref{kappa0}
shows the acceleration on the illusory horizon directly below
as seen by an observer who free-falls radially into
a Schwarzschild black hole.\cite{Hamilton:2016iid}
Since a free-faller accelerates away from the sky above,
they also see Hawking radiation from the sky above.
Figure~\ref{kappa0} shows the acceleration on the sky directly above.
The Figure recovers the usual results well outside the black hole,
that the acceleration on the illusory horizon is $1 /(4M)$,
while the acceleration on the sky is zero.
As the free-faller falls through the true horizon,
they see both accelerations increasing smoothly.
Near the singularity,
the acceleration on both illusory horizon below and sky above diverges,
tending to one third the reciprocal of the proper time $|\tau|$
left until the observer hits the singularity,
\begin{equation}
\label{kappa0s}
  \kappa \rightarrow
  {1 \over 3 |\tau|} \propto r^{-3/2}
  \quad
  \mbox{as $r \rightarrow 0$}
  \ .
\end{equation}
A more extensive calculation\cite{Hamilton:2016iid} shows that,
for a non-rotating observer (who stares fixedly in constant directions),
the perceived acceleration near the singularity tends
to the same constant~(\ref{kappa0s}) in all directions,
except for an ever-thinning band in the horizontal directions.
This points to the remarkable conclusion that the Hawking radiation
seen by a freely-falling, non-rotating observer
becomes isotropic as they approach the singular surface.

The fact that the perceived acceleration~(\ref{kappa0s}) changes in time means
that the radiation is non-thermal.
Nevertheless,
the acceleration~(\ref{kappa0s}),
which is the only timescale in the problem,
sets the natural frequency of Hawking radiation.
Encouragingly, when the same calculation is carried out in 1+1 dimensions,
the result coincides with the known exact result.\cite{Davies:1976}

\rhofig

Figure~\ref{rho}
shows the resulting quantum energy density in radiation and vacuum.
A density of one in the Figure corresponds to (in Planck units)
\begin{equation}
  {q_{\rm eff} \over 2880 \pi^2 M^4}
  \ .
\end{equation}
Radiation has a relativistic equation of state,
so its energy-momentum tensor has zero trace.
The vacuum energy then follows directly from
the trace anomaly~(\ref{traceanomaly}).
It should be cautioned that the calculation is not exact
(an exact calculation being intractable);
nevertheless there does not seem to be much wiggle room to adjust
the curves substantially.

The central conclusion of the calculation is that Hawking radiation
diverges near the singular surface,
with characteristic frequency going as $r^{-3/2}$,
and the energy density going as $r^{-6}$, the fourth power of the frequency.
The singular surface is hot.


\section{Mass inflation at the inner horizon}
\label{inflation}

\doranfig

Real astronomical black holes rotate.
The most important difference between
the Kerr geometry for a rotating black hole
and the Schwarzschild geometry for a spherical black hole
is that the Kerr geometry has an inner as well as an outer horizon.
Figure~\ref{doran} illustrates the geometry of a Kerr black hole
with spin parameter $a = 0.96$.

A major advance in understanding the interior structure of black holes was
Poisson \& Israel's\cite{Poisson:1990eh} 1990 discovery
of the mass inflation instability at the inner horizon
of a charged, spherical (Reissner-Nordstr\"om) black hole.
The inflationary instability is the nonlinear realization of
the infinite blueshift at the inner horizon first pointed out by Penrose.\cite{Penrose:1968}
Barrab\`es, Israel \& Poisson\cite{Barrabes:1990} soon generalized
Poisson \& Israel's argument to the case of a rotating black hole.

\rninnerhorizonfig

Any black hole with an inner horizon
is subject to the mass inflation instability.
Regardless of their orbital parameters,
light and matter falling into a black hole focus at the inner horizon
into one of just two directions, the principal ingoing and outgoing
null directions.
Figure~\ref{rninnerhorizon}
illustrates that an ingoing observer at the inner horizon of a
Reissner-Nordstr\"om black holes sees outgoing light and matter
focused into a divergingly blueshifted bright point.
An outgoing observer similarly sees ingoing light and matter
focused along the opposite direction.
Movies of journeys into a Reissner-Nordstr\"om black hole are at
\url{https://jila.colorado.edu/~ajsh/insidebh/rn.html}.
Unfortunately I have not yet implemented comparable movies
of Kerr black holes.

Already in his seminal note,\cite{Penrose:1968} Penrose suggested
that the diverging concentration of energy density at the inner horizon
would destabilize it,
but in subsequent work he fell short of proving as much.
Poisson \& Israel's great contribution was to calculate for the
first time the back-reaction of the counter-streaming streams
on the black hole.
They showed that the counter-streaming would drive exponential growth
of the interior mass.

\penrosemassinflationfig

Figure~\ref{penrosemassinflation} shows a Penrose diagram that illustrates
the physical reason behind the Poisson-Israel mass inflation instability.
Between the outer and inner horizons of the Kerr (or Reissner-Nordstr\"om)
geometry, the Killing time coordinate $t$
(the coordinate associated with time translation symmetry)
is a spacelike coordinate.
Infalling objects are ingoing or outgoing depending on whether they
are moving backwards or forwards in time $t$.
There is not one but two inner horizons,
an ingoing inner horizon through which ingoing objects fall,
and an outgoing inner horizon through which outgoing objects fall.
While all infalling objects are necessarily ingoing at the outer horizon,
objects on prograde\footnote{Prograde here
means having specific angular momentum greater than that of
the principal null geodesics.}
geodesics turn around and become outgoing at the inner horizon,
while objects on retrograde geodesics remain ingoing.

In their original work,
Poisson \& Israel envisaged that ingoing and outgoing streams
would be produced by a Price tail of gravitational radiation
generated by the initial collapse of a black hole
(for simplicity they used a spin-0 scalar field as a surrogate for
spin-2 gravitational waves).
Most of the literature on mass inflation
since then has focused on this same scenario.
The outcome is
a ``weak null singularity on the Cauchy horizon''
(the Cauchy horizon, the boundary of (un)predictability,
is what I have been referring to as the inner horizon).

The singularity is weak in the sense that,
although the tidal force diverges to infinity,
it does so in such a short proper time that the spacetime is scarcely distorted.
The reason for this behaviour is that the exponential timescale over which
inflation grows is proportional to the accretion rate:
smaller accretion rates imply faster inflation.\cite{Hamilton:2008zz}
The Price tail of radiation that is assumed to drive inflation
decreases to zero as a power law with time,
causing the growth rate of inflation, and with it the tidal force,
to diverge before the spacetime has had time to respond to the tidal force.

All real astronomical black holes accrete,
cosmic microwave background photons if nothing else.
Accretion fuels both ingoing (retrograde) and outgoing (prograde) streams.
It is straightforward to estimate that
when a black hole forms from stellar collapse,
the energy density in the Price tail of gravitational radiation,
which decays with time as $t^{-12}$
in its slowest decaying mode (the quadrupole),
falls below the density of the cosmic microwave background
in at most several seconds.
For black holes that continue to accrete,
the outcome is collapse, not a weak null singularity.\cite{Hamilton:2008zz}

\section{Conformally separable solutions for accreting, rotating black holes}
\label{conformallyseparable}

One of the remarkable features of the Kerr geometry
is that geodesics in it are Hamilton-Jacobi separable.\cite{Carter:1968c}
Indeed,
the hypothesis that geodesics are Hamilton-Jacobi separable
provides one way to derive the Kerr and related geometries:
Hamilton-Jacobi separability imposes conditions on the line-element
that permit the Einstein equations to be separated systematically.

It proves possible to impose the weaker condition of conformal
separability, and still separate the Einstein equations.\cite{Hamilton:2010a}
Conformal separability requires only that null geodesics, not all geodesics,
are Hamilton-Jacobi separable.
Whereas standard separability imposes that the spacetime is stationary,
conformal separability admits a time-dependent conformal factor $\rho$
in the line-element.
The conformal factor $\rho$
in a conformally separable black hole
is a product of separable (Kerr)
$\rho_{\rm s}$,
time-dependent $e^{v t}$, and inflationary $e^{-\xi}$ factors,
\begin{equation}
\label{conformalfactor}
  \rho
  =
  \rho_{\rm s}
  e^{v t - \xi}
  \ .
\end{equation}
The factor $e^{v t}$ means that the spacetime expands in a self-similar fashion.
The inflationary exponent $\xi$,
which is a function only of the self-similar radial coordinate $x$,
begins to deviate from zero only close to the inner horizon.

\inflationaryhorizonfig

Whereas stationary black holes cannot accrete,
conformal separability yields solutions for rotating
black holes that grow by accretion.
In contrast to stationary black holes,
conformally separable black holes necessarily undergo inflation
at their inner horizons.

The conformally separable black hole solutions are not exact,
but hold asymptotically in the limit $v \rightarrow 0$
of small but non-zero accretion rates.
Furthermore the solutions require a symmetric (monopole, essentially)
accretion rate.
However, what happens in the inflationary zone at any angular position
on the inner horizon depends largely on the accretion rate of
ingoing and outgoing streams on to that position.
Inflation takes place over such a small proper time that
different angular positions on the inner horizon are causally separated
from each other, and become progressively more causally separated
as inflation proceeds.
Therefore the conformally separable solutions ought to be a reliable guide
to what happens at the inner horizons of real black holes.
Full numerical relativity would be needed to test this proposition.

It might seem remarkable that conformally separable solutions for
accreting, rotating black holes exist.
The reason is that, in the limit of small accretion rates,
the geometry above the inner horizon is the usual Kerr geometry,
and then the behaviour at the inner horizon is constrained by the
fact that all matter incident on the inner horizon becomes highly
focused along either of just two directions, the principal ingoing
and outgoing null directions.
The Einstein equations enforce conservation of energy-momentum,
and the symmetry imposed by conformal separability then fixes the solution.

The solutions
Figure~\ref{inflationaryhorizon}
shows the radial ($x$) and angular ($y$) components
$T_{xx}$, $T_{xy}$, $T_{yy}$
of the energy-momenta in a conformally separable, accreting,
rotating black hole.
The solutions depend on the accretion rates
$u \mp v$
of ingoing and outgoing streams incident on the inner horizon.
The solution shown in Figure~\ref{inflationaryhorizon} is for
ingoing and outgoing accretion rates
\begin{equation}
  \mbox{in} = u - v = 0.001
  \ , \quad
  \mbox{out} = u + v = 0.003
  \ .
\end{equation}
The energy-momenta in
Figure~\ref{inflationaryhorizon}
are shown as a function of the inflationary exponent $\xi$ in the conformal
factor $\rho$, equation~(\ref{conformalfactor}).
The inflationary exponent is a function of the self-similar radial coordinate
$x$, but $x$ itself scarcely budges away from its value at the inner horizon,
whereas $\xi$ changes.
The inflationary exponent $\xi$ is almost zero well above the inner horizon,
remains small during inflation,
and then grows large as the spacetime collapses.

Figure~\ref{inflationaryhorizon}
illustrates that, for a black hole that continues to accrete,
as all astronomically realistic black holes do,
inflation gives way to collapse.\cite{Hamilton:2008zz}

The angular components of energy-momentum shown in
Figure~\ref{inflationaryhorizon}
are relative to the frame defined by the principal null directions.
The fact that the angular components are non-zero
means that the ingoing and outgoing streams in a rotating black hole
are slightly misaligned with the principal directions.

\section{Inflation gives way to BKL oscillatory collapse}
\label{bkl}

Collapse causes rotational motions to grow.
The conformally separable solution shown in
Figure~\ref{inflationaryhorizon}
fails once the spacetime has collapsed to the point that
angular energy-momentum is comparable to the radial energy-momentum.
What happens then?

During the 1970s,
Belinskii, Khalatnikov \& Lifshitz\cite{Belinskii:1970} (BKL)
(see also Belinskii\cite{Belinski:2014kba})
developed analytic arguments about the generic behaviour of spacetime
during collapse to a singularity.
They assumed that spacetime was already deep into collapse,
and they argued that under those circumstances the gradient
in the time direction would dominate gradients in spatial directions.
Thus it would be reasonable to suppose that the spacetime was
spatially homogeneous.
The eigenvectors and eigenvalues of the $3 \times 3$ spatial tetrad metric
can be thought of as describing 3 orthogonal axes with 3 different lengths.
As the spacetime evolves,
the lengths and directions of the 3 orthogonal axes evolve.

BKL showed that, in the most generic case,
spacetime collapse could be described by a series of ``Kasner epochs''
punctuated by bounces.
The Kasner line-element is\cite{Kasner:1921}
\begin{equation}
  ds^2
  =
  -\, dt^2 + a_1^2 dx_1^2 + a_2^2 dx_2^2 + a_3^2 dx_3^2
  \ ,
\end{equation}
which is a spatially homogeneous solution
of the vacuum Einstein equations provided that the scale factors $a_i$
depend as power laws with time $t$,
\begin{equation}
  a_i = | t |^{q_i}
  \ ,
\end{equation}
with exponents $q_i$ related by
\begin{equation}
\label{qi}
  q_1 + q_2 + q_3 = 1
  \ , \quad
  q_1^2 + q_2^2 + q_3^2 = 1
  \ .
\end{equation}
The time coordinate $t$ is negative, and goes to zero at the singularity,
$t \rightarrow -0$.
The relations~(\ref{qi}) between the Kasner exponents $q_a$ imply that
two axes collapse while one expands,
the mean scale factor ${\bar a} \equiv ( a_1 a_2 a_3 )^{1/3} = | t |^{1/3}$
decreasing monotonically.
During a BKL bounce,
the exponents $q_i$ change abruptly,
with one of the collapsing axes turning into expansion,
and the expanding axis going into collapse.
Generally, the bounce is accompanied by a change in the directions
of the 3 orthogonal axes.

\kninfafig

To find out what happens after the conformally separable solutions
of \S\ref{conformallyseparable} failed,
I resorted to numerical relativity.
The problem is challenging:
the numerical method must remain stable while following robustly
the physical divergence in curvature and energy-momentum.
For this purpose I developed a new
covariant Hamiltonian tetrad approach\cite{Hamilton:2016mmc}
that permits greater freedom in the choice of tetrad frame.

Because of the difficulty of the problem,
I followed the lead of BKL in assuming that spatial gradients
in horizontal directions are subdominant compared to radial gradients,
and followed the integration into the black hole along lines of
constant latitude.

Figure~\ref{kninfa}
illustrates the result of a numerical calculation
at latitude $22.5^\circ$ into a rotating black hole of spin $a = 0.96$.
The initial conditions were taken to be those of
the conformally separable solution,
and the integration was started already deep into the inflationary epoch.
Because smaller accretion rates result in larger curvatures and
energy-momenta,
it was advantageous to chose the accretion rates to be ``large,''
so that the numerical calculation would continue as long as possible
before over/underflow stuttered the calculation to a halt.
On the other hand, the accretion rates must be small enough that
the conformally separable solution provides a satisfactory approximation
to the initial conditions.
The compromise initial conditions in
Figure~\ref{kninfa}
are $u = 0.02$ and $v = 0.01$,
implying ingoing and outgoing accretion rates of
\begin{equation}
  \mbox{in} = u - v = 0.01
  \ , \quad
  \mbox{out} = u + v = 0.03
  \ .
\end{equation}

The evolution shown in 
Figure~\ref{kninfa}
is entirely characteristic of BKL collapse.
The cosmic scale factors $a_i$ evolve as power laws with time
$| t | = {\bar a}^3$
through four Kasner epochs separated by BKL bounces.
The calculation terminates when numerical over/underflow occurs.

\kninfqfig

Figure~\ref{kninfq}
shows the Kasner exponents $q_i$ measured from the evolution shown in
Figure~\ref{kninfa}.
During Kasner epochs,
the exponents indeed satisfy the Kasner condition
$\sum_i q_i^2 = 1$.

The first Kasner epoch is that of inflation,
during which the Kasner exponents are
\begin{equation}
\label{qi1}
  q_1 = 1
  \ , \quad
  q_2 = q_3 = 0
  \quad
  \mbox{inflation}
  \ .
\end{equation}
The second Kasner epoch is that of conformally separable collapse,
during which the Kasner exponents are
\begin{equation}
\label{qi2}
  q_1 = - \tfrac{1}{3}
  \ , \quad
  q_2 = q_3 = \tfrac{2}{3}
  \quad
  \mbox{collapse}
  \ .
\end{equation}
Interestingly,
the Schwarzschild geometry approximates a Kasner geometry near the
Schwarzschild singularity,
and the Kasner exponents~(\ref{qi2}) during the collapse epoch
coincide with those of the Schwarzschild geometry,
suggesting an elegant continuity between rotating and spherical black holes.

\section{The Future of Research on Black Hole Interiors}
\label{future}

What happens inside black holes is unobservable, and
inaccessible to experiment, therefore it is not very useful to carry out
theoretical research on the subject.
Does that thought resonate with you?
If so, welcome to what I believe is the principal reason for the paucity
of research on the interiors of astronomically realistic black holes.
My own view is that it is time to recognize and renounce the taboo.

The top priority areas for future research on black hole interiors are,
firstly,
Hawking radiation in the inflationary and collapse zones
of rotating black holes,
and secondly,
what happens when two rocks collide in the inflationary zone
of a rotating supermassive black hole at the centre of a galaxy.
The first problem is interesting
first to see how Hawking radiation modifies the classical evolution,
and second because
Hawking radiation gives clues about quantum gravity,
and it makes sense to look for clues near singularities.
The second problem is interesting because
this is the one place in our Universe where Nature is carrying out
collision experiments at Lorentz gamma factors of $10^{40}$ and higher,
reaching Big Bang densities and beyond.

\end{document}